\documentclass[12pt]{aastex}
\usepackage{emulateapj5}
\usepackage{epsfig}
\usepackage{graphics}

\def\etal{{\it et~al.}}
\def\ie{i.~e.}
\def\eg{e.~g.}

\def\lmxb{LMXB}
\def\msun{$M_\odot$}
\def\sce{4U~1728--34}
\def\mdot{$\dot{M}$}
\def\medd{$\dot{M}_{\rm Edd}$}

\begin{document}

\title{The Effect of the Mass Accretion Rate on the Burst Oscillations
in \sce}

\author{Lucia~M.~Franco\altaffilmark{1}}
\affil{University of Chicago, 5640 S.~Ellis Ave., Chicago IL 60637;
lucia@oddjob.uchicago.edu}
\altaffiltext{1}{Also Goddard Space Flight Center, Code 662.}


\begin{abstract}
We present a comprehensive study of the properties of nearly coherent
brightness oscillations (NCOs) in a large sample of type-I X-ray bursts
observed from low mass X-ray binary \sce. We have analyzed $\sim$545~ks
of data of this source obtained with the Rossi X-ray Timing Explorer
over a 3 year span from 1996--1999. The data contain 38 bursts, 16 of
which show NCOs. We find no NCOs present when the inferred mass
accretion rate of the system is lowest. Furthermore, we define a measure
of the strength of the oscillations and find that this strength
increases with increasing mass accretion rate. This correlation is
particularly evident within bursts detected only a few weeks apart, and
becomes less clear for bursts separated in time by several months to
years. The correlation we find for \sce\ between the NCOs and the
inferred accretion rate of the system is similar to that found for
KS~1731--260 by Muno \etal, where the NCOs are only present at
relatively high mass accretion rates.  However, unlike the case for
KS~1731--260, we find an anti-correlation in the bursts of \sce\ between
the existence of episodes of photospheric radius expansion and the
inferred mass accretion rate of the system. Moreover, we distinguish
between NCOs present in the rise and in the decay phase of the burst and
find that the bursts with NCOs only in the decay phase occur at
intermediate accretion rates, while those with NCOs in both the rise and
the decay phases are concentrated at high inferred mass accretion rates.
We discuss these results in the context of the theory of thermonuclear
bursts and propose intrinsic differences in the neutron stars in
KS~1731--260 and \sce\ as the origin of the different behaviors. 
\end{abstract}


\section{Introduction}
\label{intro}

Low mass X-ray binaries (\lmxb s), systems in which a compact object
accretes matter via Roche lobe overflow from a low mass ($<$ 1 \msun)
main sequence companion, provide a fascinating window into the physics
and dynamics of accretion disks and compact objects. When the compact 
object is a neutron star, these systems also permit detailed studies 
of the burning of the
accreted material on the neutron star surface.  Thermonuclear flashes
(called type-I X-ray bursts) occur in many \lmxb s and are caused by
explosive unstable burning of the accreted matter. The properties of
these type-I X-ray bursts are affected by the rate at which the matter
is deposited onto the neutron star \citep[see, \eg,][for an extended
review of X-ray bursts]{lvpt93}. Thus, in order to understand these
bursts, it is necessary to investigate the effect that variations in the
mass accretion rate (\mdot) onto the neutron star have on them. The
first attempts at characterizing \mdot\ in \lmxb s centered around the
persistent flux of the sources, since intuitively the larger the amount
of matter accreted, the greater the amount of X-ray flux
emitted. However, it has become clear that the persistent flux in
general may not always be a good measure of \mdot\ \citep[\eg,][]{vdkbible} 
and hence cannot be directly correlated with the burst properties.

X-ray color-color diagrams (CDs) were introduced to study subtle
spectral changes in \lmxb s \citep[see, \eg,][and references
therein]{vdkbible}. \citet{hasvdk} studied the correlated rapid X-ray
variability and spectral behavior of \lmxb s and showed that they
naturally fall into two distinct categories. These two groups are called
the atoll and Z sources, after the tracks they trace out in the CD. The
motion of the sources along these tracks is thought to be caused by
changes in the mass accretion rate \citep{hasvdk} and it has become
clear that the position of the sources on these tracks is a better
indicator of the \mdot\ of the system than the persistent flux (see
\citealt{vdkbible} for \lmxb s in general but \citealt{mariano1728} for
\sce\ in particular).

With the launch of the Rossi X-Ray Timing Explorer (RXTE) satellite,
previously unknown high frequency timing phenomena were discovered and
now offer us new views of \lmxb s \citep[see, \eg,][for an extended
review of the recently discovered timing phenomena in
\lmxb s]{vdkkhz}. In particular, nearly coherent brightness oscillations
(NCOs) with frequencies between 330 and 589~Hz were discovered in Type I
bursts from several sources
\citep{oscdisc,stroh97a,smith,zhang97,zhang98,mss99,disc1659}.  These
NCOs have extremely high coherence values \citep[$\nu / \Delta\nu >
4000$;][]{zstroh,muno}, high amplitudes \citep[up to $\sim$50\%
rms;][]{stroh97,stroh98b}, and they exhibit long term frequency
stability of up to years \citep{stroh98}. These properties strongly argue
for the rotation of the neutron star as the origin of the NCOs. In this
interpretation, during a thermonuclear burst inhomogeneities in the fuel
or in the fuel burning may arise causing ``hot spots'' on the surface of
the neutron star. These hot spots rotate with the spin frequency of the
star and periodically modulate the X-ray flux.

These NCOs can be present in the rise of the bursts, very near the burst
onset, and/or in the decay phase (or tail) of the bursts. When the NCOs
are observed in the tail, their frequency can increase by 1 -- 3~Hz,
reaching an asymptotic limit as the bursts progress \citep[see
\eg,][]{stroh98b}. It has been proposed that this frequency increase is
caused by thermal expansion of the thermonuclear burning layer on the
stellar surface. Heating from the burning expands the shell by several
tens of meters at the onset of the burst.  Due to angular momentum
conservation this layer will slow down causing the observed frequency to
decrease. During the decay of the burst, the burning layer settles back
on the neutron star surface and spins up causing the frequencies to
increase again. In this picture, the asymptotic limit represents the
true neutron star spin frequency \citep{stroh97}.  The atoll source
4U~1636--53 has shown evidence for a subharmonic to the NCO frequency,
indicating the possible existence of two, nearly antipodal hot spots
\citep[perhaps at the magnetic poles;][]{miller99}. This detection of
the subharmonic raises issues about how the burning front propagates on
the surface of the neutron star and how the fuel can be concentrated in
two antipodal regions.

The burst oscillations in the atoll source KS~1731--260 have been
recently studied by \citet{muno}. They reported the presence of a
correlation between the inferred \mdot\ of the system and the presence
of NCOs in the bursts. They find that the bursts from KS~1731--260 can
be separated into two classes: the so called ``fast'' bursts, which
occur at relatively high \mdot, exhibit NCOs, and show episodes of
photospheric radius expansion (PRE, the expansion of the neutron star
photosphere occurring when the burst luminosity is higher than the
Eddington limit), and the ``slow'' bursts, which occur at lower \mdot,
do not exhibit NCOs, and do not show PRE episodes.  Studies like this
require a large sample of bursts observed over a considerable range of
accretion rates. Such data bases have so far only been obtained for the
persistent sources because the outbursts of the X-ray transients are
rare and when they do occurr only a limited number of X-ray bursts are
seen. In this paper, we will concentrate on the persistent source \sce.

The \lmxb\ \sce\ is a prolific source of type I X-ray bursts \citep{basinska} and
one of the best studied \lmxb s. It has been classified as an atoll source
\citep{hasvdk,mariano} and it exhibits most of the behaviors seen in
those systems. It was the first neutron star \lmxb\ to show twin kilohertz
quasi-periodic oscillations (kHz QPOs) in the persistent emission {\it
and\/} the first to show NCOs \citep{oscdisc}.  The kHz QPOs in \sce\ have
been studied by several groups
\citep{oscdisc,fordkhz,mariano,disalvokhz} and the NCOs by
\citet{oscdisc}, \citet{stroh97,stroh98}, and \citet{zstroh}.  In this
paper we present a study of the behavior of NCOs in a set of 38 type-I
X-ray bursts from \sce, observed over a wide range of mass accretion
rates. We found that \sce\ is similar to KS~1731-260 in that the NCOs 
are only present in bursts which occur at high \mdot\ \citep{muno}. 
However, contrary to the case of KS~1731--260, we find an
anti-correlation between the presence of PRE episodes in the bursts and
the inferred mass accretion rate of the system. We also find a more
complicated behaviour with \mdot\ for the NCOs in the rise and in the
decay phases of bursts.

This paper is organized as follows: in \S\ref{obs} we describe in detail
the observations being used for this study. In \S\ref{analysis} we
explain the methods used to characterize the persistent emission and the
burst properties, and the techniques used to study the oscillations. In
\S\ref{cdsect} we discuss the spectral states (\ie, the CD) of the
source. In \S\ref{osccorr} we discuss the behavior of the NCOs and study
them in correlation with other burst characteristics and the spectral
state of the source.  Finally in \S\ref{discussion} we discuss the
implications of these correlations for our current understanding of
\sce\ in particular and for thermonuclear burning on the surface of
neutron stars in general.

\section{The observations}
\label{obs}

The source \sce\ was observed with the Proportional Counter Array (PCA) on board
RXTE on many occasions between 1996 February 15 and March 1 (hereafter
referred to as observation set AO1), 1997 September 19 and October 1
(AO2), and 1998 September 30 and 1999 January 18 (AO3). The actual
observation dates are listed in Tables \ref{ao1}, \ref{ao2} and
\ref{ao3}, along with the main properties of all 65 observations in our
sample. In total $\sim$545~ks of on-source data were obtained. The AO1
observations were obtained with epoch 1 gain settings for the PCA
instrument, while AO2 and AO3 observations correspond to epoch 3 gain
settings. Hereafter, when we refer to an observation we mean those data
that can be identified by an unique observation identification number or
ObsID (see Tables~\ref{ao1}, \ref{ao2}, and \ref{ao3}).  All 5 PCA detectors were
on during all observations except for 3.6 ks during the observation with
ObsID 20083-01-01-020 and 3.6 ks during 20083-01-03-020. No bursts were
detected during those two intervals, so we have excluded those data from
our study.  The analysis of portions of our data have already been
published in the literature. \citet{oscdisc} reported the discovery of
NCOs using burst 5 in our sample and \citet{stroh97} presented a
temporal and spectral analysis of bursts 3, 4 and 5, and reported the
existence of NCOs in bursts 4 and 5. \citet{mariano} published a CD of AO1
and a portion of AO2 data \citep[see also][]{disalvokhz}.

A variety of data modes was available for each observation set in
addition to the Standard 1 and Standard 2 modes. Because the same data
modes were 
\begin{table*}[t]
\parbox{3.6in}{\caption{\label{ao1} AO1 Observations of \sce\ (RXTE PCA Epoch 1)}}\hspace{0.1in}\parbox{3.6in}{\caption{\label{ao2} AO2 Observations of \sce\ (RXTE PCA Epoch 3)}}\\
\setlength{\tabcolsep}{1.2mm}
\begin{minipage}{7in}
\begin{minipage}{3.4in}
\begin{tabular*}{3.5in}{llcccrcc}
\hline
\hline
\rule[-0mm]{0mm}{0pt}\\
Obs. ID & Date & \multicolumn{2}{l}{Start} & End & $T_{\mathnormal{exp}}$\footnote{Total on-source time per ObsID} & 
$F_{\mathnormal{avg}}$\footnote{Average flux in units of $10^{-9}$ erg s$^{-1}$
cm$^{-2}$ with uncertainties of 1\% -- 3\% (90\% confidence interval).} & Burst\\
(10073-01) & 1996 & \multicolumn{2}{l}{Time} & Time & (ks) &  & \\
\rule[0mm]{0mm}{0mm}\\
\hline
\rule[0mm]{0mm}{0pt}\\
01-000 & Feb 15 & 11:51 & & 18:49 & 11.8 & 4.1 &  1 \\
01-00  & Feb 15 & 18:49 & & 00:02 & 13.1 & 4.2 &  2 \\
02-000 & Feb 16 & 00:02 & & 06:22 & 13.1 & 4.4 &  3 \\
02-00  & Feb 16 & 06:22 & & 10:14 & 10.1 & 4.7 &  4,5 \\
03-000 & Feb 16 & 15:48 & & 22:11 & 12.3 & 4.7 &  6 \\
03-00  & Feb 16 & 22:11 & & 01:19 &  7.7 & 4.8 &   \\
04-000 & Feb 18 & 11:08 & & 18:49 & 13.4 & 3.4 &  7 \\
04-00  & Feb 18 & 18:49 & & 23:49 & 10.5 & 3.2 &  8 \\
06-000 & Feb 22 & 11:32 & & 18:53 &  8.9 & 2.0 &   \\
06-00  & Feb 22 & 18:53 & & 00:02 & 10.1 & 3.3 &  9 \\
07-000 & Feb 23 & 21:16 & & 05:16 & 18.2 & 2.3 &   \\
07-00  & Feb 24 & 05:16 & & 06:45 &  3.2 & 2.3 &  10 \\
08-000 & Feb 24 & 17:51 & & 01:26 & 15.5 & 2.4 &  11 \\
08-00  & Feb 25 & 01:26 & & 05:11 &  8.5 & 2.4 &   \\
09-000 & Feb 25 & 20:45 & & 04:45 & 17.9 & 2.5 &  12 \\
10-00  & Feb 29 & 23:09 & & 00:02 &  1.1 & 2.8 &   \\
10-01  & Mar 01 & 00:02 & & 05:45 & 13.3 & 2.8 &   \\
\rule[5mm]{0mm}{0mm}\\
\hline
\vspace{-0.4in}
\end{tabular*}

\end{minipage}\hspace{0.3in}
\begin{minipage}{3.4in}
\begin{tabular*}{3.5in}{llcccrcc}
\hline
\hline
\rule[-0mm]{0mm}{0pt}\\
Obs. ID & Date & \multicolumn{2}{l}{Start} & End & $T_{\mathnormal{exp}}$\footnote{Total on-source time per ObsID} & 
$F_{\mathnormal{avg}}$\footnote{Average flux in units of $10^{-9}$ erg s$^{-1}$
cm$^{-2}$ with uncertainties of 1\% -- 3\% (90\% confidence interval).} & Burst\\
(20083-01) & 1997 & \multicolumn{2}{l}{Time} & Time & (ks) &  & \\
\rule[0mm]{0mm}{0mm}\\
\hline
\rule[0mm]{0mm}{0pt}\\
01-00  & Sep 19 & 05:52 & & 10:40 & 11.3 & 4.8 &  \\
01-01  & Sep 19 & 12:29 & & 15:02 &  6.6 & 4.8 & 13 \\
01-020 & Sep 20 & 07:32 & & 14:03 & 14.7 & 4.1 & 14 \\
01-02  & Sep 20 & 14:03 & & 16:29 &  6.2 & 4.8 &  \\
02-01  & Sep 21 & 15:43 & & 21:05 & 13.8 & 5.0 & 15,16 \\
02-000 & Sep 22 & 05:59 & & 13:59 &  3.3 & 4.4 & 17 \\
03-01  & Sep 23 & 23:50 & & 00:41 &  1.3 & 4.8 &  \\
03-000 & Sep 24 & 09:14 & & 17:14 & 18.9 & 5.7 &  \\
03-00  & Sep 24 & 17:14 & & 18:18 &  3.7 & 5.4 &  \\
04-00  & Sep 26 & 12:29 & & 19:55 & 17.9 & 3.9 & 18,19 \\
04-01  & Sep 27 & 09:18 & & 16:43 & 18.1 & 3.2 & 20,21 \\
03-020 & Sep 30 & 04:33 & & 10:39 & 13.2 & 3.9 &  \\
03-02  & Sep 30 & 10:39 & & 14:05 &  7.1 & 3.9 &  \\
04-020 & Oct 01 & 06:09 & & 14:09 & 17.5 & 3.6 &  \\
04-02  & Oct 01 & 14:09 & & 15:18 &  3.5 & 3.7 &  \\
\rule[0mm]{0mm}{0mm}\\
\hline
\vspace{-0.4in}
\end{tabular*}

\end{minipage}
\end{minipage}
\end{table*}
\noindent not always available across observation sets, we describe here
the extant modes and how they were used to extract the quantities of
interest to this study. The AO1 data were obtained in one burst trigger
mode (TLA\_1s\_10\_249\_1s\_5000\_F), two burst catcher modes
(CB\_125us\_1M\_0\_249\_H with 122~$\mu$s time resolution in one energy
channel; CB\_8ms\_64M\_0\_249\_H with 8 ms resolution in 64 channels)
and one single bit mode (SB\_125us\_0\_249\_1s with 122~$\mu$s
resolution in one channel). The burst catcher modes provide data
typically for 1~s before the burst onset and for as long as the count
rate remains above the trigger level. In our case the length of data
ranges from 13 to 16~s for these modes. For AO2 we had the same burst
trigger and catcher modes available as during AO1, however, instead of
the single bit mode we now had one event mode (E\_16us\_64M\_0\_1s with
16~$\mu$s resolution and 64 channels) available. The data for AO3 were
obtained in three single bit modes (SB\_125us\_0\_13\_1s,
SB\_125us\_14\_17\_1s and SB\_125us\_18\_23\_1s all with 122~$\mu$s
resolution and one channel) and one event mode (E\_125us\_64M\_24\_1s
with 122~$\mu$s time resolution and 64 energy channels). When combined,
all four data modes for AO3 data cover the complete RXTE energy range.
To generate the CD we extracted light curves from the Standard 2 data
in 4 energy ranges to calculate the colors. We also used the Standard
2 data to generate spectra of the persistent emission and extract
average fluxes. The power density spectra of the persistent emission
were calculated from the single bit mode data for AO1, the event mode
data for AO2, and the data of the combined modes for AO3. These same
data modes were also used to search for NCOs in the bursts. Finally,
the time series of the burst spectra were obtained from the event mode
for AO2, the combined data modes for AO3, and the 8~ms resolution
burst catcher mode for AO1 (hence there is at most 16~s of time
resolved spectra for each burst in AO1).\\

\section{The Analysis}
\label{analysis}

To characterize the persistent emission, we calculated average fluxes
for each observation. The fluxes were obtained by fitting (using Xspec;
\citealt{xspec}) a cut-off power law spectrum to the
background-subtracted Standard 2 data. A Gaussian line around 6~keV was
added in order to obtain acceptable fits. Since we are only interested
in the flux of the persistent emission, we did not investigate the
significance of this feature further \citep[but see,
\eg,][]{disalvo}. The resulting averaged fluxes in the 3--18~keV range
are listed in Tables \ref{ao1}, \ref{ao2}, and \ref{ao3}. In order to
\begin{table*}[t]
\parbox{3.6in}{\caption{\label{ao3} AO3 Observations of \sce\ (RXTE PCA Epoch 3)}}\hspace{0.1in}\parbox{3.6in}{\caption{Properties of Bursts from 4U 1728-34\label{btable}}}\\
\setlength{\tabcolsep}{1.2mm}
\begin{minipage}{7in}
\begin{minipage}{3.4in}
\vspace{-1.5in}
\begin{tabular*}{3.4in}{llcccrcc}
\hline
\hline
\rule[-0mm]{0mm}{0pt}\\
Obs. ID & Date & \multicolumn{2}{l}{Start} & End & $T_{\mathnormal{exp}}$\footnote{Total on-source time per ObsID} & 
$F_{\mathnormal{avg}}$\footnote{Average flux in units of $10^{-9}$ erg s$^{-1}$
cm$^{-2}$ with uncertainties of 1\% -- 3\% (90\% confidence interval).} & Burst\\
(30042-03) & 1998-99 & \multicolumn{2}{l}{Time} & Time & (ks) &  & \\
\rule[0mm]{0mm}{0mm}\\
\hline
\rule[0mm]{0mm}{0pt}\\
01-00  & Sep 30 & 07:11 & & 11:52 & 10.4 & 4.4 & 22 \\
02-00  & Oct 07 & 05:05 & & 09:49 & 11.1 & 5.9 &  \\
01-01  & Oct 23 & 22:35 & & 23:08 &  1.9 & 3.8 &  \\
02-01  & Oct 24 & 00:15 & & 01:09 &  2.5 & 3.8 &  \\
03-01  & Oct 24 & 01:47 & & 03:46 &  4.8 & 3.8 & 23 \\
01-02  & Oct 26 & 20:57 & & 21:34 &  1.7 & 3.9 &  \\
01-03  & Oct 26 & 22:33 & & 23:10 &  2.1 & 3.9 &  \\
04-00  & Oct 27 & 00:09 & & 03:19 &  6.9 & 3.9 &  \\
01-04  & Oct 29 & 01:44 & & 02:43 &  3.5 & 4.0 &  \\
05-00  & Oct 29 & 03:20 & & 06:37 &  7.0 & 3.9 &  \\
06-00  & Nov 01 & 02:30 & & 07:26 &  9.8 & 4.0 & 24,25 \\
07-01  & Nov 02 & 22:29 & & 23:30 &  3.6 & 4.0 & 26 \\
07-00  & Nov 03 & 00:05 & & 03:18 &  7.4 & 4.0 & 27 \\
08-00  & Nov 05 & 04:00 & & 07:29 &  6.7 & 4.1 &  \\
09-00  & Nov 10 & 13:17 & & 13:55 &  2.2 & 4.4 &  \\
09-02  & Nov 10 & 16:31 & & 17:07 &  2.1 & 4.4 &  \\
10-00  & Nov 10 & 18:26 & & 20:19 &  4.9 & 4.4 & 28 \\
10-01  & Nov 10 & 21:07 & & 21:54 &  2.8 & 4.4 &  \\
11-00  & Nov 10 & 22:26 & & 03:16 & 10.8 & 4.3 & 29 \\
12-00  & Nov 11 & 14:50 & & 21:52 & 17.9 & 4.3 & 30,31 \\
13-00  & Nov 11 & 22:26 & & 03:13 & 10.4 & 4.4 & 32 \\
14-02  & Nov 16 & 06:34 & & 07:00 &  1.5 & 5.2 &  \\
14-01  & Nov 16 & 08:14 & & 09:08 &  3.2 & 5.2 &  \\
14-00  & Nov 16 & 09:52 & & 10:47 &  3.2 & 5.1 & 33 \\
15-00  & Nov 16 & 13:09 & & 19:15 & 14.9 & 5.1 & 34 \\
16-00  & Nov 17 & 00:12 & & 07:36 & 17.1 & 4.5 & 35 \\
17-00  & Nov 17 & 13:02 & & 14:00 &  3.4 & 4.0 & 36 \\
18-00  & Jan 16 & 07:51 & & 11:52 &  9.9 & 3.1 &  \\
19-01  & Jan 17 & 04:50 & & 05:34 &  2.6 & 3.1 &  \\
19-00  & Jan 17 & 06:11 & & 10:10 & 10.0 & 3.2 & 37 \\
20-00  & Jan 18 & 22:39 & & 01:18 &  5.4 & 3.3 & 38 \\
\rule[0mm]{0mm}{0mm}\\
\hline
\vspace{-0.4in}
\end{tabular*}

\end{minipage}\hspace{0.2in}
\begin{minipage}{3.4in}
\begin{tabular*}{3.8in}{ccccrccc}
\hline
\hline
\rule[-0mm]{0mm}{0pt}\\
No. & Date & Start & $F_{\rm pk}$\footnote{Measured peak flux in the $3-18$ keV range in units of $10^{-8}$ erg s$^{-1}$ cm$^{-2}$.} & $F_{\rm BB}$\footnote{Peak bolometric flux from best-fit black body model, in units of $10^{-8}$ erg s$^{-1}$ cm$^{-2}$.} & PRE & rise & tail\\
  &  & Time UT &  &  &  & NCO & NCO\\
\rule[0mm]{0mm}{0mm}\\
\hline
\rule[0mm]{0mm}{0pt}\\
 1 & 1996 Feb 15 & 17:58:11 & 7.3 &  8.9(8) & Y & - & Y \\
 2 & 1996 Feb 15 & 21:10:21 & 7.9 &  9.4(8) & Y & - & Y \\
 3 & 1996 Feb 16 & 03:57:10 & 4.5 &  5.4(6) & - & - & - \\
 4 & 1996 Feb 16 & 06:51:08 & 7.2 & 11.(1)  & - & Y & Y \\
 5 & 1996 Feb 16 & 10:00:44 & 7.2 & 11.0(7) & - & Y & Y \\
 6 & 1996 Feb 16 & 19:27:11 & 7.8 & 10.(1)  & - & - & Y \\
 7 & 1996 Feb 18 & 17:31:50 & 7.7 & 11.(1)  & Y & - & Y \\
 8 & 1996 Feb 18 & 21:28:50 & 8.1 & 12.(1)  & Y & - & - \\
 9 & 1996 Feb 22 & 23:09:12 & 9.2 & 14.(1)  & Y & - & - \\
10 & 1996 Feb 24 & 05:46:24 & 8.7 & 14.(1)  & Y & - & - \\
11 & 1996 Feb 24 & 17:51:49 & 8.8 & 13.(1)  & Y & - & - \\
12 & 1996 Feb 25 & 23:17:07 & 8.8 & 13.(1)  & Y & - & - \\
\\					                    
13 & 1997 Sep 19 & 12:32:56 & 5.3 &  6.4(7) & - & Y & - \\
14 & 1997 Sep 20 & 10:08:50 & 3.4 &  4.0(5) & - & Y & Y \\
15 & 1997 Sep 21 & 15:45:28 & 4.8 &  5.7(6) & - & Y & - \\
16 & 1997 Sep 21 & 18:11:04 & 4.6 &  5.5(6) & - & Y & Y \\
17 & 1997 Sep 22 & 06:42:51 & 3.0 &  3.6(5) & - & Y & Y \\
18 & 1997 Sep 26 & 14:44:09 & 4.9 &  5.8(7) & - & Y & Y \\
19 & 1997 Sep 26 & 17:29:47 & 4.7 &  5.5(6) & - & Y & Y \\
20 & 1997 Sep 27 & 11:19:06 & 8.5 & 10.2(8) & Y & - & Y \\
21 & 1997 Sep 27 & 15:54:03 & 7.9 &  9.4(8) & Y & - & Y \\
\\					                    
22 & 1998 Sep 30 & 10:08:54 & 9.4 &  11.(1) & Y & - & - \\
23 & 1998 Oct 24 & 02:32:28 & 9.3 &  12.(1) & - & - & - \\
24 & 1998 Nov 01 & 03:47:55 & 9.8 &  12.(1) & - & - & - \\
25 & 1998 Nov 01 & 06:59:11 & 9.9 &  12.(1) & Y & - & - \\
26 & 1998 Nov 02 & 22:42:59 & 9.7 &  12.(1) & Y & - & - \\
27 & 1998 Nov 03 & 02:02:24 & 9.5 &  11.(1) & Y & - & - \\
28 & 1998 Nov 10 & 19:33:39 & 8.3 &  10.(1) & - & - & - \\
29 & 1998 Nov 11 & 00:37:33 & 9.2 &  11.(1) & Y & - & - \\
30 & 1998 Nov 11 & 16:20:37 & 8.9 &  11.(1) & Y & - & - \\
31 & 1998 Nov 11 & 19:32:04 & 8.8 &  11.(1) & Y & - & - \\
32 & 1998 Nov 12 & 00:19:54 & 8.6 &  10.(1) & - & - & - \\
33 & 1998 Nov 16 & 10:10:25 & 8.4 &  10.(1) & Y & - & - \\
34 & 1998 Nov 16 & 16:09:03 & 8.6 &  10.(1) & Y & - & Y \\
35 & 1998 Nov 17 & 07:30:33 & 8.0 &  10.(1) & - & - & - \\
36 & 1998 Nov 17 & 13:44:06 & 8.8 &  11.(1) & Y & - & - \\
37 & 1999 Jan 17 & 06:12:01 & 8.6 &  10.(1) & Y & - & - \\
38 & 1999 Jan 18 & 23:46:33 & 9.6 &  12.(1) & Y & - & - \\
\rule[0mm]{0mm}{0mm}\\
\hline
\vspace{-0.4in}
\end{tabular*}

\end{minipage}
\end{minipage}
\end{table*}
verify the reliability of our fluxes we fit other spectral models
(3-component models: a comptonized spectrum plus either a multi-color
disk blackbody or a plain black body plus a Gaussian line around 6~keV;
see, \eg, \citealt{disalvo}). We find that these alternative models give
the same average flux of the persistent emission within the quoted
errors.

\begin{figure*}[t]
\vspace{1.0in}
\begin{minipage}{7.2in}
\begin{minipage}{3.5in}
\epsfig{file=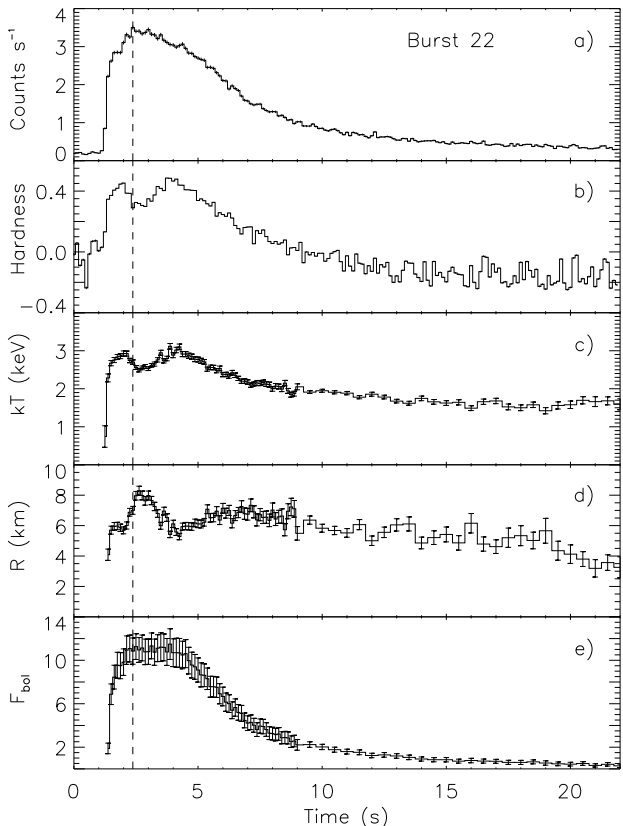,width=3.4in,bbllx=54pt,bblly=73pt,bburx=558pt,bbury=576pt}\\
\parbox{3.5in}{
\caption{\label{pre}Typical burst (burst 22) with photospheric radius
expansion. ({\it a\/}) The count rate (in units of $10^4$), 
({\it b\/}) the logarithm of the ratio of counts in the 9--18~keV range to 
that in the 3--4~keV range, ({\it c\/}) the black body temperature, 
({\it d\/}) the inferred radius of the emitting region, and ({\it e\/}) the
bolometric flux (in units of $10^{-8}$ erg cm$^{-2}$ s$^{-1}$) as a function
of the time in the burst. The dashed vertical line marks the time at
which the burst reached peak count rate.  }}
\end{minipage}\hspace{0.2in}
\begin{minipage}{3.42in}
\vspace{-1.2in}
\noindent perform a standard spectrum analysis of these bursts using
Xspec. We find that the spectra of X-ray bursts are well fitted by a
simple black body model \citep{swank77}. As a background spectrum we
used the spectrum of an interval of 10~s of the persistent emission
before the burst onset.\footnotemark[3] While the true background
spectrum during a burst may differ from that of the persistent emission
before the burst \citep{lvp86} this procedure will suffice to establish
the existence of PRE episodes.

\hspace{0.15in}From the spectral fits we extracted time series of temperature, radius
of the emitting region, and measured flux. The peak measured fluxes are
listed in Table \ref{btable}. From the best fit parameters and the black
body assumption we can calculate a peak bolometric flux using
$F_{\rm BB} = 4 \pi R^2 \sigma T^4$, which we also list in Table
\ref{btable}.  We verified the existence of PRE episodes by calculating
hardness ratios as a function of time during the burst. In
Figure~\ref{pre} we show a typical burst with a PRE episode. The drop in
the hardness ratio in ({\it b\/}) indicates the beginning of the radius
expansion phase. This behaviour is confirmed by a simultaneous decrease
in the black body temperature ({\it c\/}) and increase in the inferred
radius of the emitting region ({\it d\/}) while the bolometric flux
remains constant ({\it e\/}).

\hspace{0.15in}To search for oscillations in the bursts, we employ the $Z_n^2$
statistic described by \citet[see also \citealt{znsdetail}]{zstroh}. The
main reasons for this choice are that it does not require binning and
its probability density is $\chi^2$ distributed even for small numbers
of events (X-ray photons in our case). Since the oscillations are highly
sinusoidal, we consider only $Z_1^2$ \citep[see][]{zstroh}. We generate
data segments of 2~s in duration with centroids separated by 0.25~s
(hence they are not independent), and search for power in the vicinity
of the previously found oscillation frequency of 363~Hz
\citep{oscdisc}. We set the detection \hfill threshold at $Z_1^2 = 14$. This is
\hfill equivalent \hfill to a
\vspace{-0.6in}
\end{minipage}
\end{minipage}
\end{figure*}

To generate the CD, for AO1 we define the soft color as the logarithm
of the ratio of the count rate in the energy range 2.9--4.5~keV to the
count rate in the range 4.5--6.3~keV. Similarly we define the hard
color as the logarithm of the ratio of the count rate in the range
6.3--9.0~keV to that in the range 9.0--18.2~keV. For AO2 and AO3 the
energy ranges are slightly different: 3.0--4.4~keV to 4.4--6.2~keV for
the soft color and 6.2--9.1~keV to 9.1--18.2~keV for the hard
color. The colors were calculated from background-subtracted light
curves from the Standard 2 data and hence have 16~s time
resolution. However, to reduce the errors on the colors we averaged
them in 256~s intervals.

\begin{figure*}
\epsfig{file=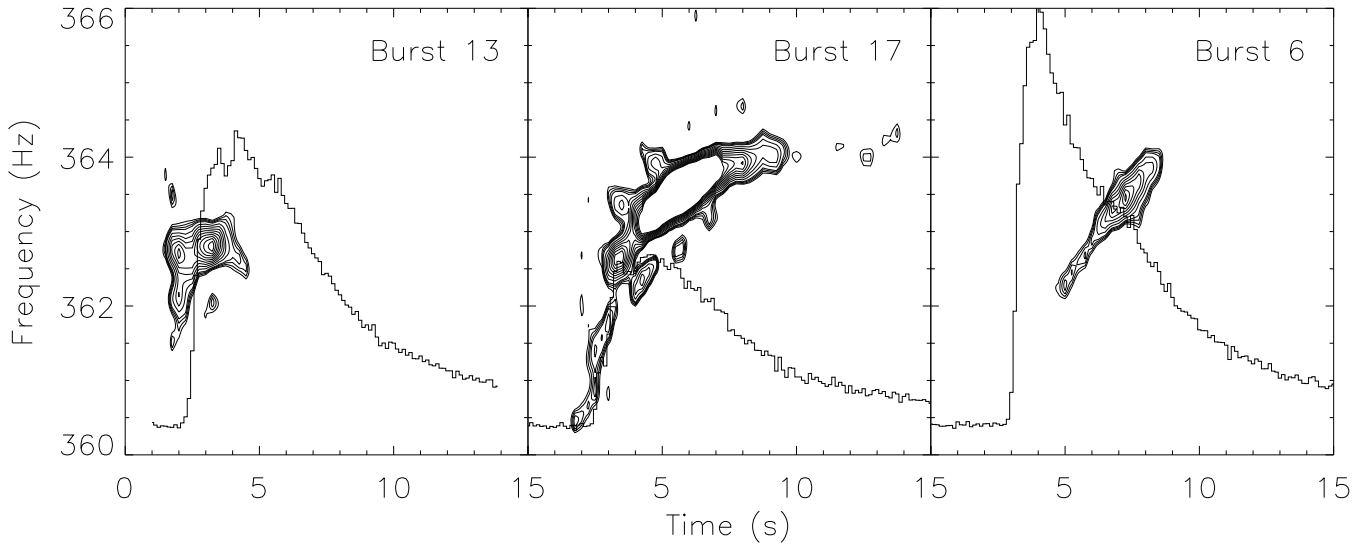,width=7.0in,bbllx=76pt,bblly=84pt,bburx=545pt,bbury=270pt}
\caption{\label{zns}Representative dynamical power spectra for bursts
with oscillations only in the rise of the burst (left), in both the rise
and the decay portions of the burst (middle) and only in the decay phase
of the burst (right). The first contour plotted corresponds to $Z_1^2 =
14$. The contours increase in steps of 2 until $Z_1^2=20$ and then
they increase in steps of 5 until $Z_1^2=65$. In the figures, the burst
profiles are overplotted in the same arbitrary units.}
\end{figure*}

We searched for bursts by generating light curves of the Standard 1
data with 0.125~s resolution. We found 38 bursts in our data
set. We have numbered the bursts according to time and have listed
their main properties in Table \ref{btable}. The start time of the
burst is determined from the first time bin in the
background-subtracted light curve with count rate above 1000 counts
s$^{-1}$.  We characterize the bursts by obtaining time resolved
series of spectra. We extracted dead-time corrected spectra of 0.125 s
duration for a time interval of 20~s during the bursts.\footnote{Except
for AO1 observations where the only data mode available with high
enough time and energy resolution was a burst catcher mode with only
up to 16~s per burst.} We\footnotetext[3]{Except again for AO1
observations where only up to 2~s of data was available prior to burst
onset.}

\noindent single trial significance of a little better than $1 \times
10^{-3}$. Since pulsations are known to exist in this source we do not
need to search a wide frequency space and we therefore have few
effective trials.  Using this criteria we found that 16 out of our
sample of 38 bursts exhibit NCOs. Typical examples of bursts which show
NCOs are shown in Figure~\ref{zns}.  We established a measure of the
strength of the oscillations by integrating the RMS amplitude of the
oscillations over their duration in the burst, ignoring the wings of the
power features which fall below our detection threshold.

\section{The Color-Color Diagram of \sce}
\label{cdsect}

\begin{figure*}[t]
\epsfig{file=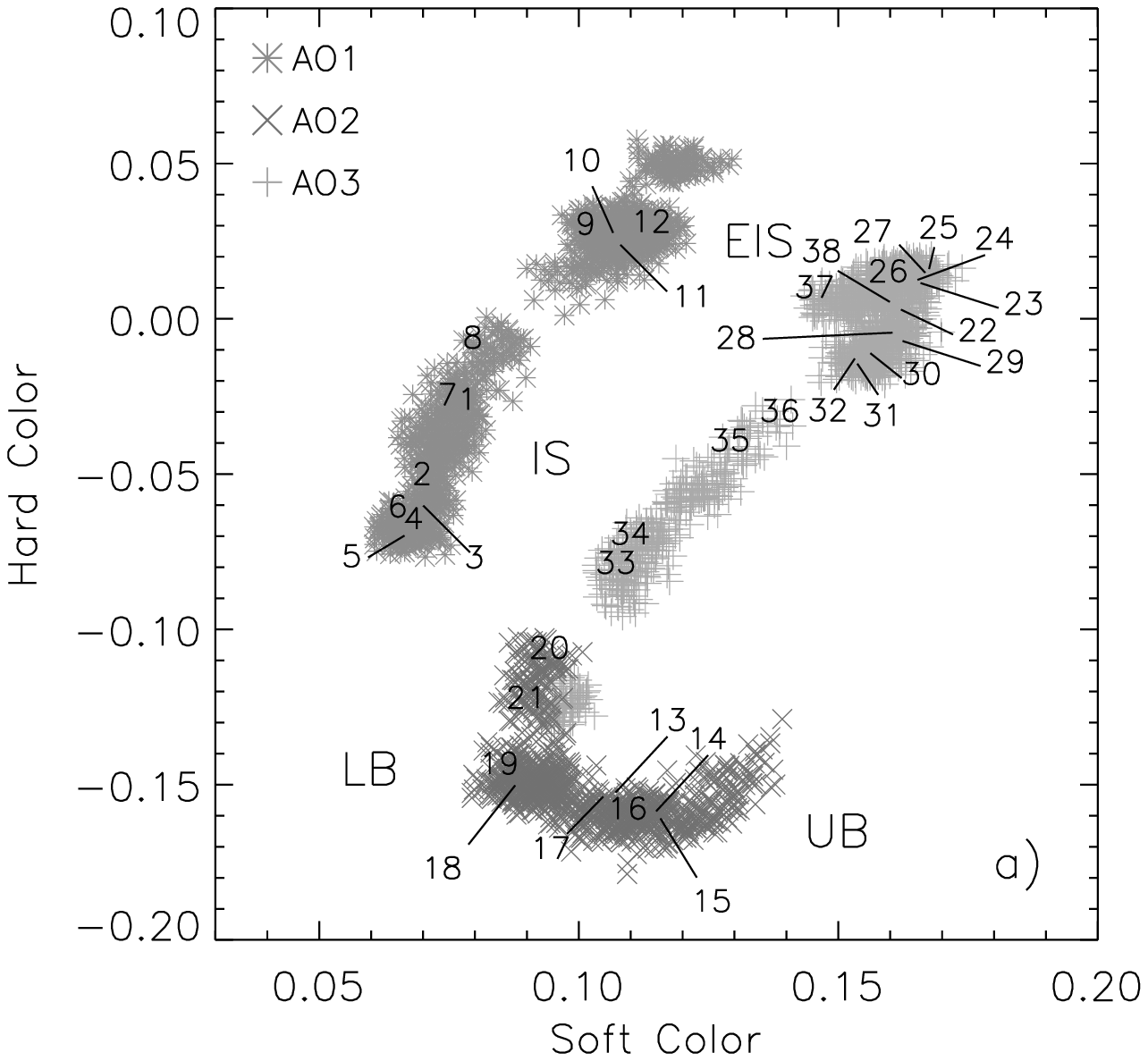,width=3.5in,bbllx=54pt,bblly=360pt,bburx=430pt,bbury=710pt}\epsfig{file=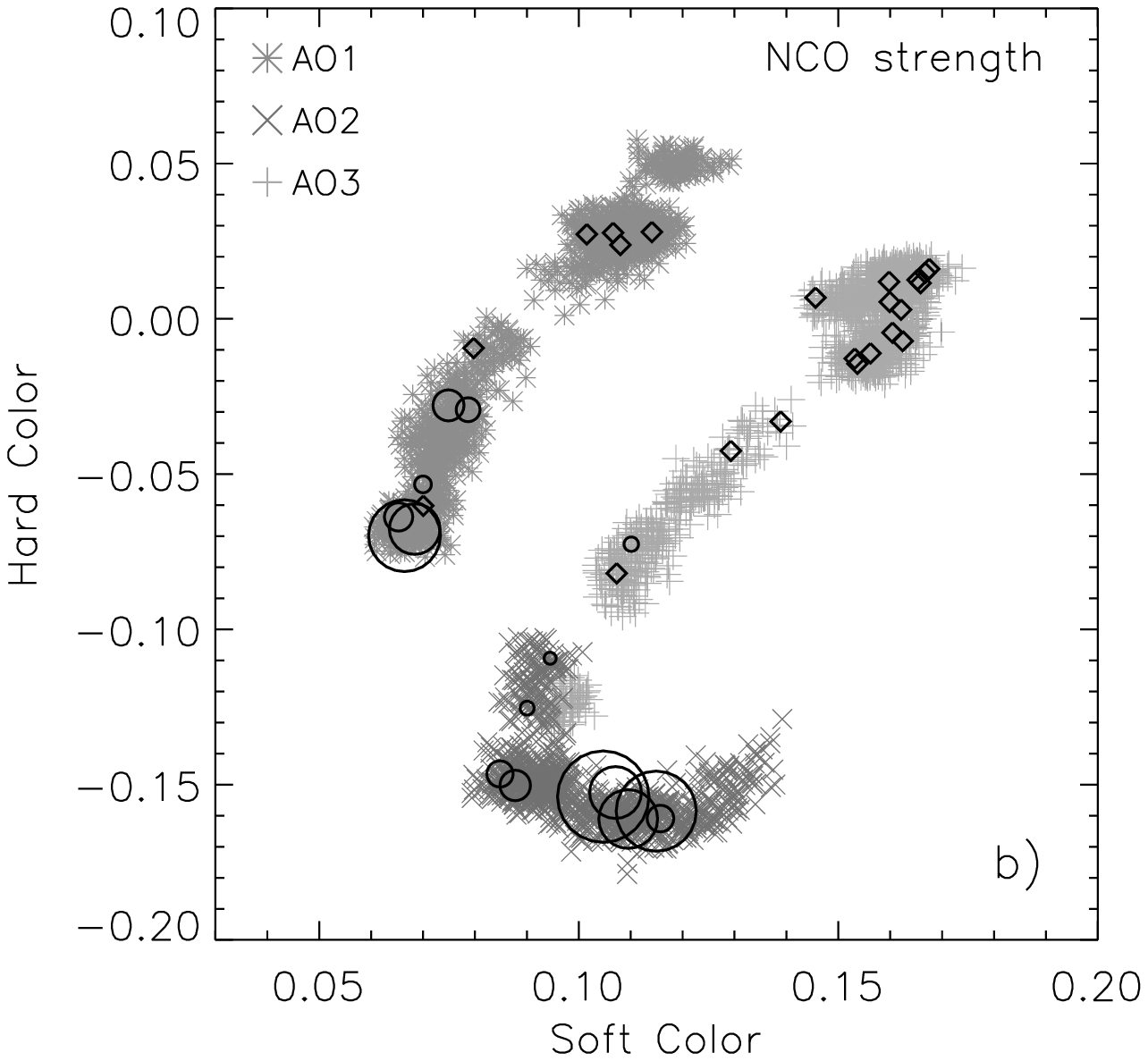,width=3.5in,bbllx=54pt,bblly=360pt,bburx=430pt,bbury=710pt}\\
\epsfig{file=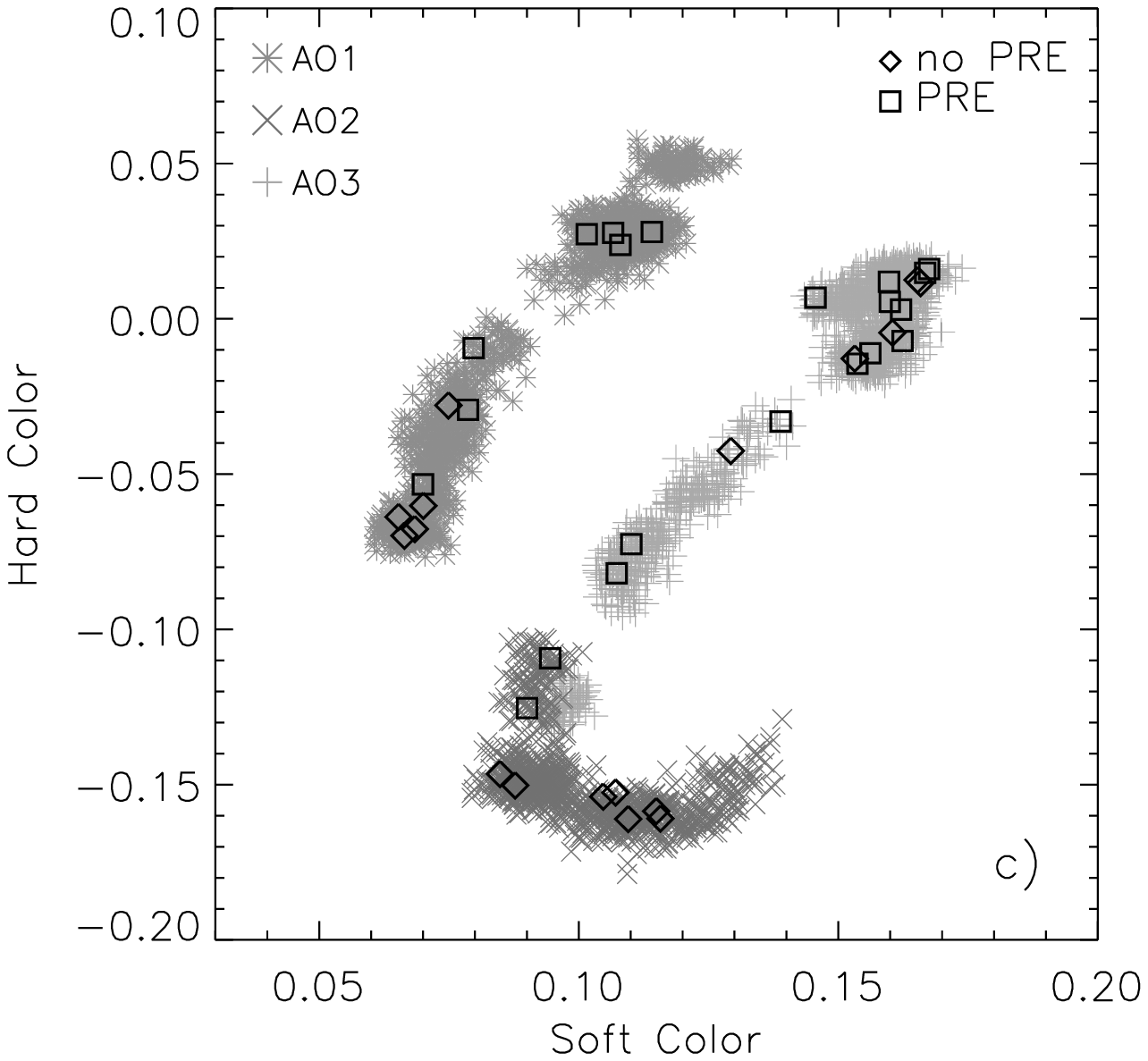,width=3.5in,bbllx=54pt,bblly=360pt,bburx=430pt,bbury=710pt}\epsfig{file=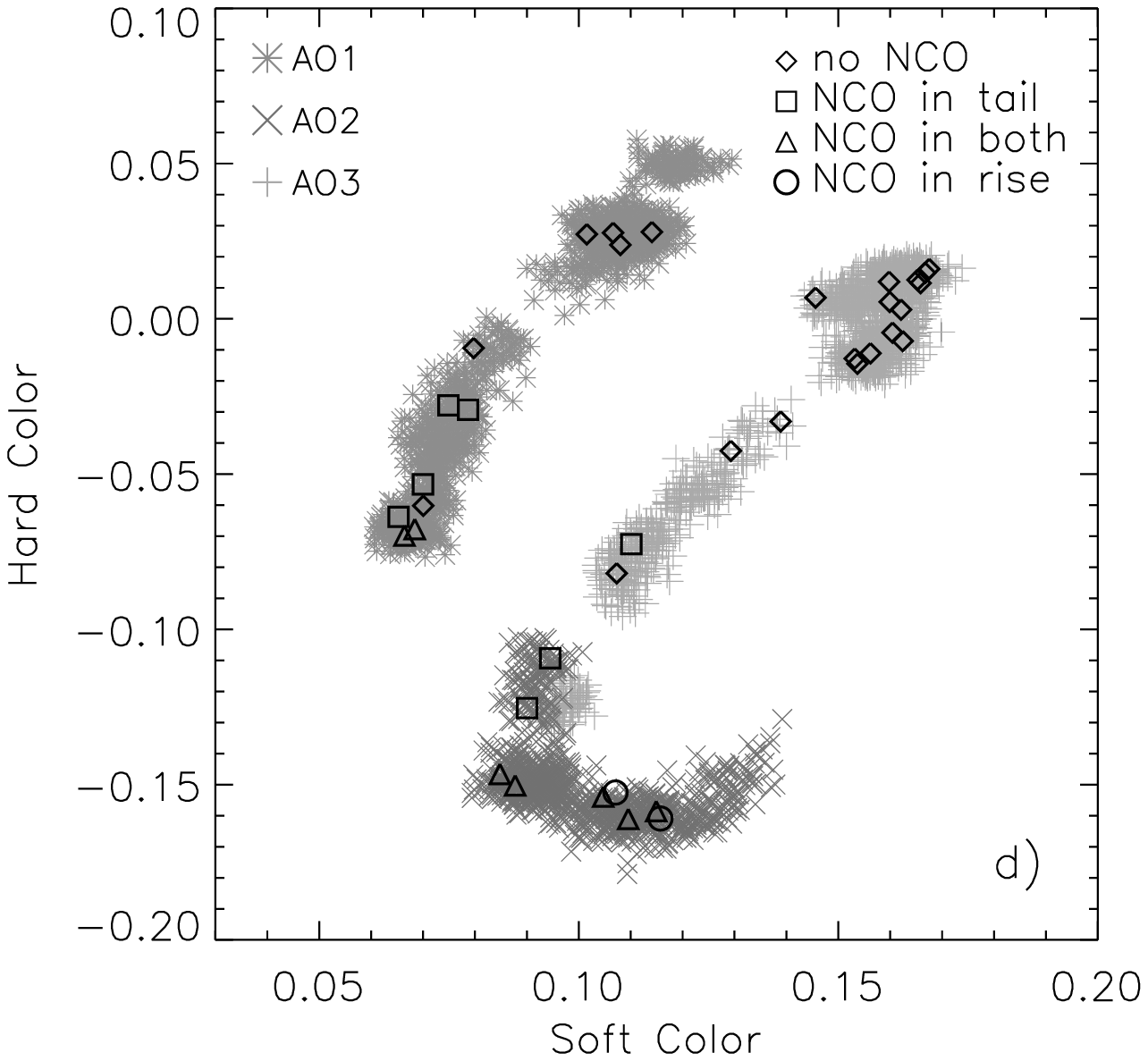,width=3.5in,bbllx=54pt,bblly=360pt,bburx=430pt,bbury=710pt}
\caption{\label{cd}{\bf a)} Color-color diagram for \sce\ showing a
typical atoll track and counterclockwise from upper right, the extreme
island state (EIS), the island state (IS), the lower banana branch (LB),
and the upper banana branch (UB). The stars correspond to data obtained
with epoch 1 instrument gain settings (AO1 data), while the exes
(AO2) and crosses (AO3) belong to epoch 2. The shift in the tracks between
AO1 and AO3 is due to these gain differences and is not intrinsic to the
source. Each point represents 256~s of data. The numbers indicate the
state of the source prior to the corresponding burst.
{\bf b)} The same
diagram now showing the presence and the strength of the nearly coherent
brightness oscillations (NCOs) in each burst. The diamonds are bursts
without NCOs. The circles correspond to bursts with NCOs and the size of
the circle is proportional to the strength of the NCOs (see text).
{\bf c)} The same diagram showing the
presence of episodes of photospheric radius expansion (PRE). Bursts that
exhibit PRE episodes are shown with squares, while those that do not are
shown with diamonds.  
{\bf d)} The same diagram showing the
presence of the NCOs only in the cooling tail of the bursts ({\it
squares\/}), both in the tail and the rise of the bursts ({\it
triangles\/}), and only in the rise of the bursts ({\it circles\/}).
Diamonds mark the bursts without detectable NCOs.  }
\end{figure*}

The CD of \sce\ in Figure~\ref{cd}{\it a} shows a track typical of atoll
sources (\citealt{hasvdk}, see also \citealt{mariano1728}).  It is
important to note that the shift in the CD of the track for AO1 with
respect to the tracks of AO2 and AO3 is not an intrinsic behavior of
the source but due to the difference in gain settings. Our
observations of \sce\ sampled both the island state and the banana
branch, though not during a single observation set. Both AO1 and AO3
exhibit the complete island state track, from the extreme island state
at the upper right of the CD, believed to be the state with the lowest
mass accretion rate, to just before the island state--banana branch
vertex (hereafter simply referred to as 'the vertex'). Only during our
AO2 observations was \sce\ found in the banana branch. This
observation set covers the entire banana branch, from the lower banana
branch on the left corner to the upper banana branch on the bottom
rightmost portion of the track, where the mass accretion rate is
thought to be the highest. Our CD looks very similar in shape to those
published previously for \sce\ \citep{mariano,piraino,disalvokhz}. Our
island state appears somewhat longer, but this difference is due to
the different choice of colors and disappears when similar colors are
used.

To determine the position of the source in the CD prior to each burst,
we determine the color of the source from an interval of 320~s before
the burst, ending 16~s prior to the onset of the burst. The results can
be seen in Figure~\ref{cd}{\it a} where the numbers
indicate the spectral state of the source just prior to the occurrence
of each burst. From the figure it is clear that we detected bursts
throughout almost the entire track, from the extreme island state to
the upper banana branch although not in the uppermost part of the
upper banana branch. However, it is not yet clear whether this is due
to a real lack of bursts at the highest $\dot{M}$ or due to the
limited amount of data so far obtained when \sce\ was at such high
accretion rates.

\begin{figure*}[t]
\epsfig{file=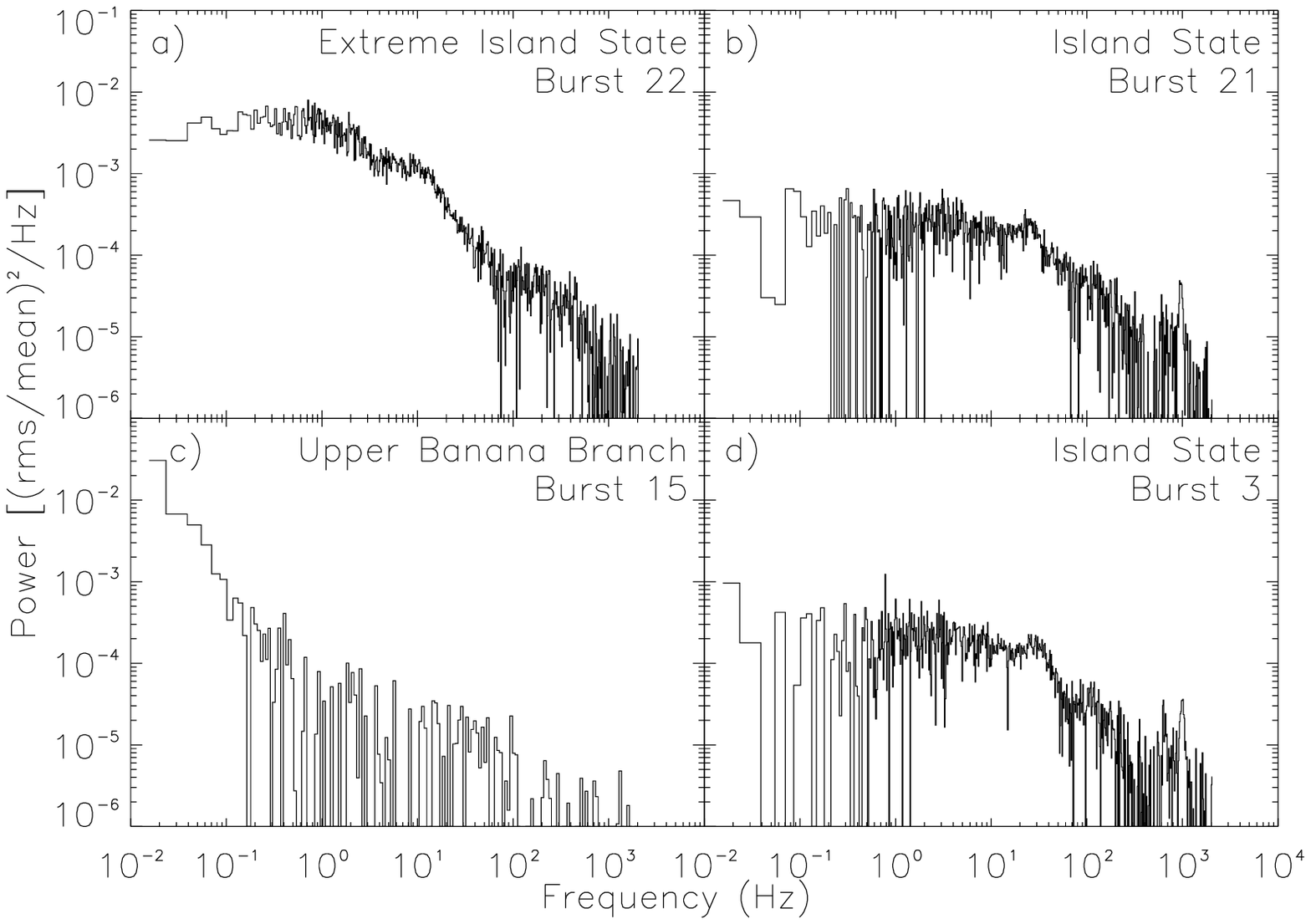,width=6.0in,bbllx=54pt,bblly=360pt,bburx=558pt,bbury=720pt}
\caption{\label{pds}Representative power density spectra from bursts
occurring when \sce\ was in the extreme island state ({\it a}), vertex
between island state and banana branch ({\it b, d\/}), and upper banana branch
({\it c}).}
\end{figure*}

To verify the spectral state of the source prior to each burst,
average power density spectra (PDS) were obtained by integrating over
the entire length of the continuous data train prior to each burst.
Figure~\ref{pds}{\it a\/} shows the PDS of the source prior to burst
22, which is typical

\noindent of the source when it is in the extreme
island state. This PDS clearly shows band-limited noise at frequencies
above 100~Hz, with no high frequency QPOs, a bump around 10~Hz, a flat
spectrum at low frequencies, and a break around 1~Hz, typical for the
island state \citep[see, \eg,][]{vdkbible}.  At the vertex, \eg,
prior to burst 21, the source PDS (Fig.~\ref{pds}{\it b\/}) shows a kHz
QPO and a break around 20~Hz typical of the lower banana branch in
transition to the island state. Figure~\ref{pds}{\it c\/} shows the PDS
for the source prior to burst 15, when \sce\ is in the upper banana
branch. The PDS shows a typical featureless power law noise
component. As mentioned above, the difference in gain settings between
epoch 1 and 3 causes a shift in the island state tracks between AO1
and AO3. To see where the AO1 track might join the AO2 track if there
were no gain differences, we show in Figure~\ref{pds}{\it d\/} the PDS
of the source prior to burst 3. This PDS is very similar to that of
burst 21 (Fig.~\ref{pds}{\it b\/}), indicating a transition between the
island state and the lower banana branch.

\section{The Behavior of Burst Oscillations}
\label{osccorr}

\subsection{The NCOs and the Inferred Mass Accretion Rate}

As shown in Figure~\ref{favg}, we find no correlation between the
average flux of the persistent emission prior to the bursts and the
existence or the strength of the NCOs. However, if the presence and
possibly the strength of the NCOs is correlated with the mass accretion
rate, as was the case in KS~1731--260 (Muno \etal\ 2000), then the lack
of correlation is not surprising, because, as already mentioned in
\S~\ref{intro}, the flux is not a robust tracer of the mass accretion
rate. Instead, a better measure of \mdot\ is the position of the source
on the atoll track in the CD.

In Figure~\ref{cd}{\it b}, we plot all bursts in the CD; with diamonds
the bursts that do not show NCOs, and with circles those that do. For bursts
with oscillations the size of the circles is proportional to the strength
of their NCOs as defined in \S~\ref{analysis}. From this figure it is 
clear that bursts 
with NCOs occur only on the bottom part of the diagram where the accretion rate
is thought to be highest, \ie, near the vertex and on the banana
branch.  Close inspection of this figure reveals that the strength of
the NCOs generally increases with decreasing hard color, \ie, the NCOs
become stronger as the inferred mass accretion rate of the system
increases. This relation is especially evident for bursts within a
single observation set, and hence separated in time by only a few days
to a few weeks, but it is not so clear for bursts from different
observation sets, separated in time by up to years (\ie, between the
AO1 and the AO2 data sets).

\subsection{The NCOs and the Burst Flux}

\begin{figure*}[t]
\begin{minipage}{7.2in}
\begin{minipage}{3.5in}
\epsfig{file=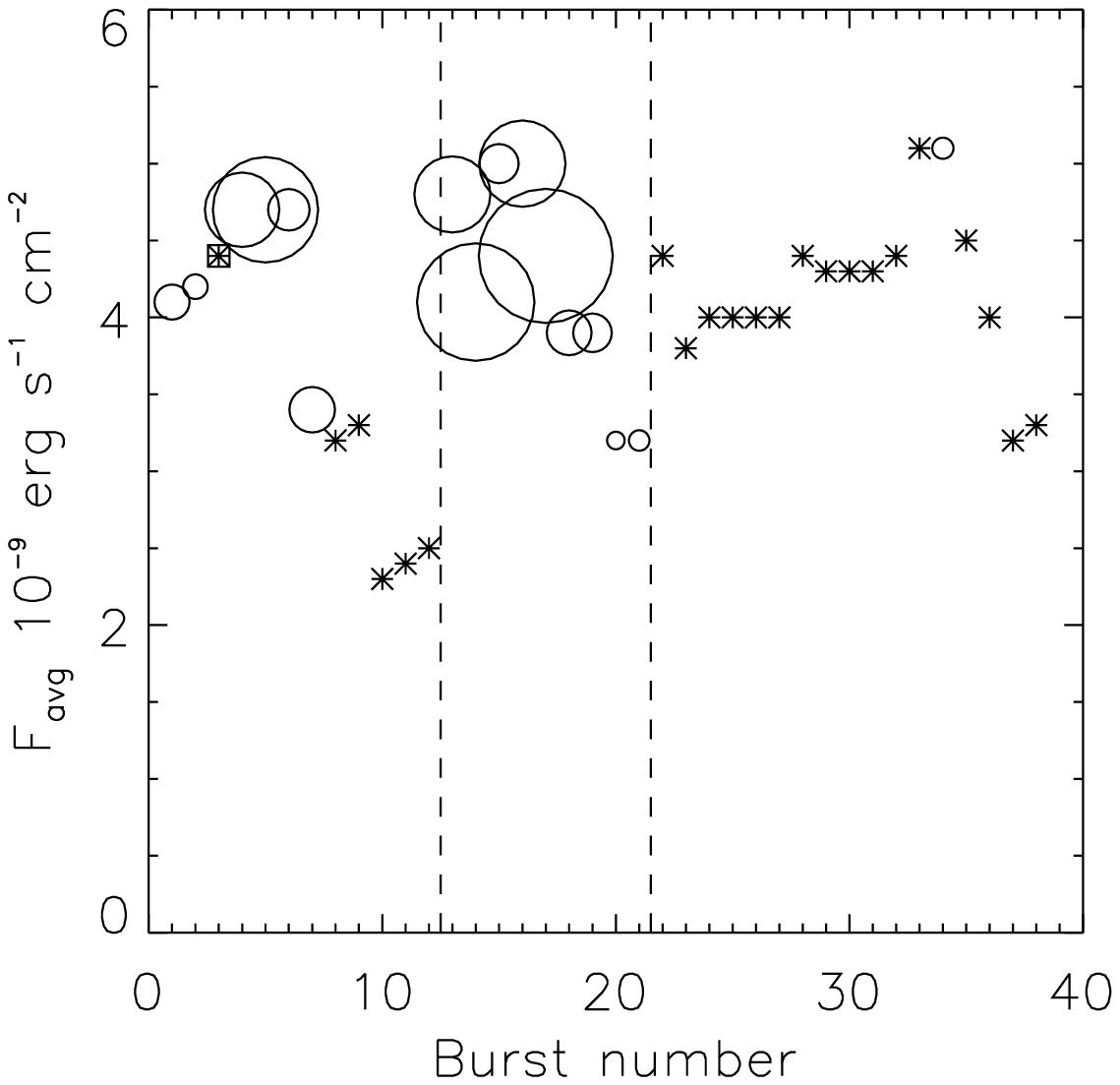,width=3.2in,bbllx=106pt,bblly=376pt,bburx=441pt,bbury=692pt}
\parbox{3.5in}{
\caption{\label{favg} Average persistent emission fluxes obtained from a
power law fit to the data prior to each burst. The dashed lines separate the
bursts during AO2 observations from those of AO1 and AO3. Bursts
without NCOs are shown with asterisks, while bursts with NCOs are
shown with circles whose size is proportional to the strength of the
oscillations (see text). Burst 3 which exhibits an unusual light curve
is marked with a square.}}
\end{minipage}\hspace{0.2in}
\begin{minipage}{3.5in}
\epsfig{file=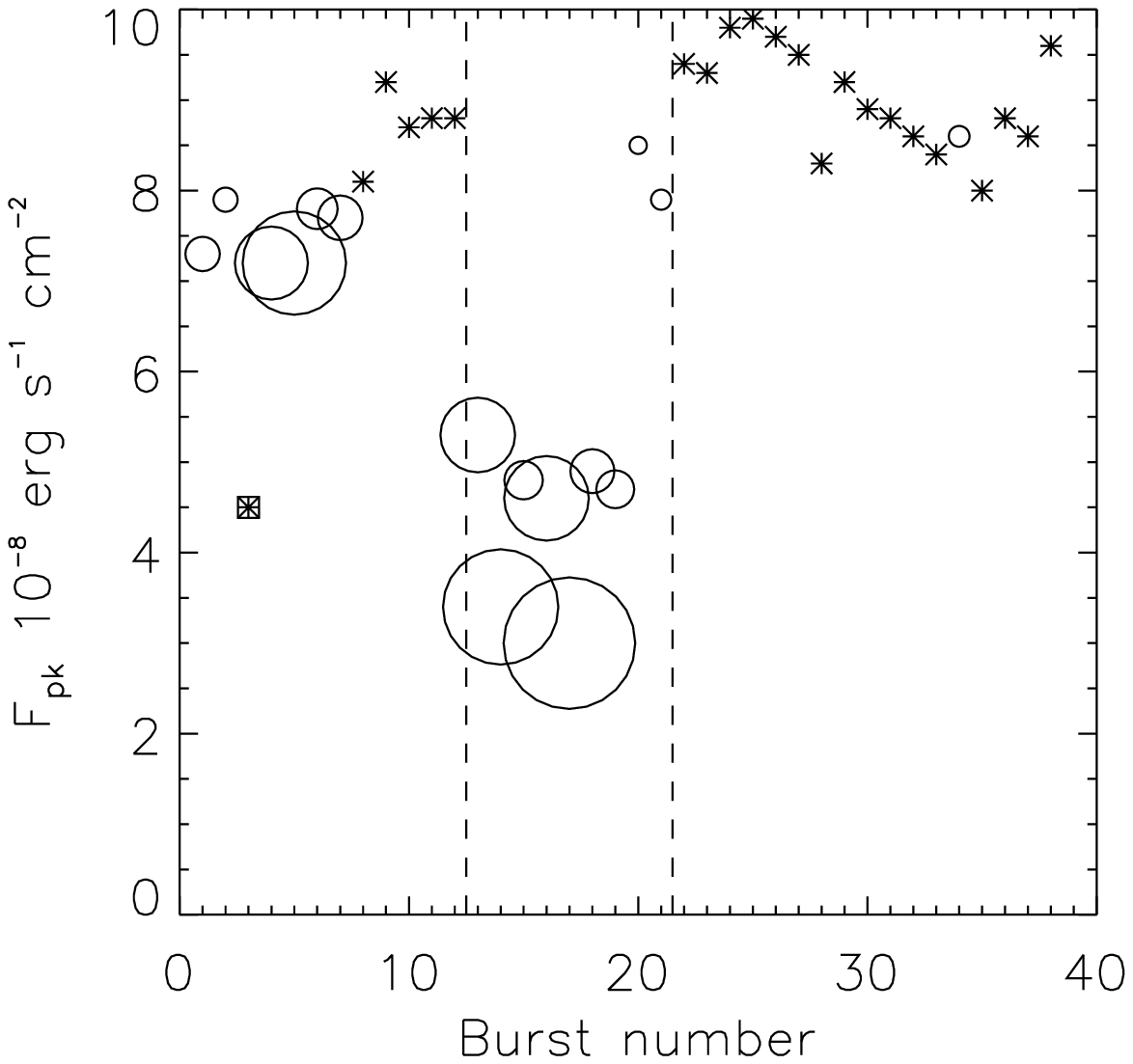,width=3.2in,bbllx=106pt,bblly=376pt,bburx=441pt,bbury=692pt}
\parbox{3.5in}{
\caption{\label{fpk} Peak burst fluxes obtained from a black body fit
to running spectra of 0.125~s duration. The dashed lines separate the
bursts during AO2 observations from those of AO1 and AO3. Bursts
without NCOs are shown with asterisks, while bursts with NCOs are
shown with circles whose size is proportional to the strength of the
oscillations (see text). Burst 3 which exhibits an unusual light curve
is marked with a square.}}
\end{minipage}
\end{minipage}
\end{figure*}

In Figure~\ref{fpk} we show the peak burst fluxes for the bursts in
our sample. Bursts without NCOs are marked with asterisks, while
bursts with NCOs are shown with circles whose radius is proportional
to the strength of the oscillations. Burst 3 which shows an unusual
count rate profile is marked with a square. The first trend evident
from Figure~\ref{fpk} is that the bursts in our AO2 observations
(except for bursts 20 and 21) have considerably lower peak fluxes than
those in our AO1 and AO3 observations. We also observe that the bursts
without NCOs have on average higher peak fluxes. Furthermore, among
the bursts with oscillations in a single observation set (AO1 {\it or\/}
AO2), the dimmest bursts also have the strongest oscillations. This
correlation does not hold when we consider bursts from AO1 and AO2
together.

\subsection{The NCOs and Bursts with Photospheric Radius Expansion}

In their study of KS~1731-260, \citet{muno} found that all the bursts
with PRE episodes occur at high mass accretion rates and they all show
NCOs. They go further to distinguish between oscillations occurring
during the rise of the burst and those occurring during the cooling
tail of the burst. They note that oscillations in the rise of bursts
are not correlated with PRE episodes since burst 7 in their sample
exhibits such NCOs but not a PRE episode. However, their data suggests
that all bursts with NCOs in the decay portion of the burst do exhibit
PRE episodes.

A careful look at Figure~\ref{cd}{\it c} reveals a different picture
for \sce . Bursts with PRE episodes (shown with squares) occur
throughout the entire island state and at the vertex, but not in the
banana branch itself. This is precisely the opposite of the behavior
in KS~1731-260 where all bursts with PRE episodes lie in the banana
branch, at the highest \mdot . We also note that all 7 bursts from our
sample that lie in the banana branch (but not bursts 20 and 21 which
lie near the vertex) have very low peak measured fluxes (see
Fig.~\ref{fpk}) while all the PRE bursts in the KS~1731--260 study are
``fast'' bursts which show high peak fluxes.

The behavior of the bursts in \sce\ is richer still. 
\citet{muno} hinted at a possible difference between the NCOs
observed during the rise of bursts and those observed in the cooling
tails of the bursts. If we make this distinction we find in our sample
2 bursts with NCOs {\it only in the rise\/}, 7 bursts with NCOs in {\it
both the rise and the tail\/}, and 7 bursts with NCOs {\it only in the
tail\/}. Representative bursts of each of these categories are shown in
Figure~\ref{zns}. In Figure~\ref{cd}{\it d} we can see that these
distinctions in the NCOs are also borne out on the CD. Bursts with
NCOs only in the decay tail of the burst are seen (marked with
squares) at intermediate \mdot, in the island state and the
vertex. Those bursts with NCOs in both the rise and the tail of the
bursts are found (marked with triangles) only at high \mdot, in the
banana branch. The two bursts with NCOs only in the rise of the burst
are shown with circles and they appear in the upper banana branch, at
the highest inferred \mdot. However, with only two such bursts it is
difficult to determine whether or not their location only in the UB
branch is significant. The bursts without NCOs are shown with
diamonds.

\section{Discussion}
\label{discussion}

We have searched three years of archival RXTE data of the neutron star
\lmxb\ and atoll source \sce\ for type-I X-ray bursts. We have found 38
bursts, 16 of which show nearly coherent brightness oscillations (NCOs).
The presence or absence of the NCOs in the bursts is not well correlated
with the persistent flux of the source, but it is very well correlated
with the position of the source on the atoll track in the CD. The NCOs
are only present in the bursts that occur when the source is near the
island state-banana branch vertex or on the banana branch.  It is
thought \citep{hasvdk} that the position of the source in the atoll
track is determined by the mass accretion rate. If true, then the
presence of the NCOs is very well correlated with the inferred mass
accretion rate and the lack of any correlation with the persistent X-ray
flux is further proof that the persistent flux is not a good measure of
the accretion rate of the system.

A similar study for KS~1731--260 by \citet{muno}, although with less
bursts (9 bursts, 5 with NCOs), also found that the NCOs are only
present in bursts which occur at relatively high mass accretion
rates. However, \citet{muno} also report that the bursts of KS~1731--260
which have episodes of photospheric radius expansion (PRE) occur at
these high accretion rates and that they all exhibit NCOs. Such a
correlation can be excluded for \sce. When we considered bursts that
exhibit (PRE) episodes, we discovered a
strong anti-correlation with inferred mass accretion rate: bursts with
PRE episodes occur only when the source is at its lowest and
intermediate mass accretion rates and not at its highest accretion
rates.  Also, the presence of NCOs during the bursts is not correlated
with the presence of PRE episodes: both bursts with and without NCOs can
exhibit periods of PRE. Yet another difference we found is that the
bursts which occured during relatively high mass accretion rates were
the dimmest in our sample, while the bursts in KS~1731--260 were bright
when the accretion rate in that system was high \citep{muno}.  It is
unclear why the positions of bursts in the CD are very similar with
respect to the existence of NCOs in both \sce~and KS~1731--260, but
other burst properties differ so significantly with CD position (and
thus with the inferred mass accretion rate).

One clue to these differences may be found in the theory of
thermonuclear bursts. Atoll sources accrete mass at rates in the range
0.01 \medd $<$ \mdot $<$ 0.1 \medd\ \citep{hf82}. In this regime, the
model of \citet{mdotmodel} for thermonuclear bursts \citep[see
also][]{bildsten97} predicts two main behaviors dominated by the
accretion rate. At relatively low mass accretion rates, the hydrogen on
the neutron star surface burns stably forming a layer of pure helium. As
more matter accumulates this layer is compressed and heated until it
ignites unstably, producing a pure helium burst. As the mass accretion
rate increases, hydrogen is accreted faster than it can be burned and
when the helium flash ignities it does so in a mixed hydrogen/helium
environment.  For atoll sources, bursts occurring in the island state at
low inferred mass accretion rates would then be the result of pure
helium flashes, with the fraction of hydrogen fuel increasing as the
source progresses toward the banana branch.

Because of the relative timescales for energy release of the reactions
involved (the CNO cycle for hydrogen burning is limited by the rate of
$\beta$-decay and thus much slower than the triple alpha process for
helium burning), bursts with high fractions of helium are expected to
release their energy faster and reach higher peak fluxes. This
behavior seems borne out in the bursts from \sce. From Table
\ref{btable} and Figure~\ref{cd}{\it a} we can see that in our AO1
observations, the bursts in the extreme island state (bursts 9--12)
have higher peak fluxes than the rest of the bursts in that
observation set. Similarly within the AO3 observations the
bursts at the lowest \mdot\ (22--27 and 38) also have the highest peak
fluxes among their group. It is natural then that episodes of
photospheric radius expansion are anti-correlated with mass accretion
rate; the lower the \mdot, the higher the helium fraction, the faster
the energy release and the brighter the bursts, making it more likely
that they will reach the Eddington limit and expand the photosphere.

In the case of KS~1731--260, \citet{muno} argue that the inverse
behavior of this source with respect to the theory's predictions may
be explained by considering not the overall mass accretion rate but
the rate of mass accretion local to the fuel surface. If the area over
which the matter is being accreted increases with mass accretion rate,
the local accretion rate may in fact decrease, causing the inverse
behavior \citep{bildsten2000}. The question then becomes, why does the
area over which the accreted matter is distributed increase with
increased \mdot\ in KS~1731--260 but not in \sce? The answer may lie in
the intrinsic differences of the neutron stars in each of these
systems. For example, a significant difference in the magnetic field
(\ie, its strength and orientation) of the neutron star in
\sce~compared to that in KS~1731--260 may inhibit the motion of matter
from the equator to higher latitudes.  Alternatively, if the frequency
of the NCOs is indeed the true spin frequency of the neutron star,
then KS~1731--260 with an NCO asymptotic frequency of 524~Hz is
spinning nearly 45\% faster than \sce\ (with an asymptotic NCO
frequency of 363~Hz). This may reduce the effective gravity at the
equator of the neutron star in KS~1731--260 and allow more significant
spread of the accreted matter.

If KS~1731--260 and \sce\ are so different with respect to their burst
behavior, why then are burst oscillations not observed at low inferred
mass accretion rates in either of these sources? Evidently the
mechanism producing the oscillations is strongly influenced by the
mass accretion rate of the system.  Since the NCOs are likely closely
tied to the properties of the nuclear burning it is very possible that
we are seeing the direct influence of mass accretion rate on the
burning behavior. \citet{muno} point to recent work by \citet{cumming}
which indicates that modulations of the burst emission are more likely
observable during helium-rich bursts, which they argue occur in the
banana branch for KS~1731--260 where the NCOs are observed in this
source. This cannot be the case for \sce\ as we have argued that the
bursts with NCOs in \sce\ have in fact high fractions of hydrogen
mixed in which causes them to be dim. Further work on possible ``hot
spot'' or other alternative mechanisms will greatly illuminate this
issue.

Although it is likely that a low mass accretion rate directly inhibits
the oscillation production mechanism, we explore the possibility that
these oscillations are in fact present during the burst but that
environmental changes (due to the differences in \mdot) cause the
oscillations to become undetectable at low \mdot. Recently, the X-ray
eclipsing source X 1658--298 was found to also exhibit NCOs during
five bursts \citep{disc1659}. Several more bursts were found which did
not show these oscillations.  However, there is strong evidence that
the absence of NCOs in those bursts was caused by environmental
changes in the system.  In X 1658--298, the oscillations were
undetectable or only marginally detectable at times when the source
exhibited clear episodes of dipping behavior, making it likely that
the mechanism behind the dips causes the oscillations to become
undetectable \citep{disc1659}.  The dips can be understood as an
obscuration of the inner system by matter periodically coming into the
line of sight. This same matter then also attenuates the oscillations
making them undetectable. Although no X-ray dips have been observed
for both \sce~and KS~1731--260, it is possible that the absence of
NCOs at low mass accretion rates in these sources might also be due to
changes in the environment.  It is known that when \mdot\ decreases in
all atoll sources, the X-ray spectrum becomes significantly harder and
follows a power law spectrum \citep[see, \eg,][for a recent study of
the hard X-ray spectrum of several atoll sources]{barret}. It has been
postulated that this power law spectrum is due to a comptonizing
medium (\eg, a corona) around the inner system.  If such a medium is
indeed present at low \mdot\ but not at high \mdot, it might be
responsible for smearing out the oscillations making them unobservable
at low \mdot. Similar mechanisms have also been introduced to explain
the absence of the expected coherent pulsations in the persistent
X-ray emission of neutron star \lmxb s.

However, this interpretation is not without problems. It has been shown
that the only \lmxb\ for which coherent millisecond oscillations have so
far been found during the persistent emission (SAX J1808.4--3658;
\citealt{1808disc}) was in the island state (as judged both from the
aperiodic rapid variability and the power law spectrum;
\citealt{1808aperiod,gilfanov}) during the observations which led to its
discovery as a millisecond X-ray pulsar. So, if a comptonizing medium is
present in \sce\ and KS~1731--260 which blurs the NCOs during the type-I
X-ray bursts, why are the pulsations in SAX J1808.4--3658 not
attenuated? This suggests that the mass accretion rate has a more direct
influence on the production mechanism of the NCOs.

Such a direct influence of \mdot\ on the NCOs is also suggested by the
fact that for seven bursts in \sce\ the oscillations were visible during
both the rise and the decay phases of the burst, while seven bursts
show oscillations only in the decay phase and two bursts show
oscillations during the rise but not during the tail of the burst.  We
find that in \sce\ bursts with NCOs only during the tail occur at
intermediate mass accretion rates, while bursts with NCOs in both the
rise and the decay occur at high mass accretion rates. Our data
further suggest that bursts with NCOs only in the rise occur at the
highest mass accretion rates. Clearly, the exact moment when the NCOs
will be present during the bursts is determined by the mass accretion
rate, which strongly suggests that the NCOs are indeed heavily
influenced by the mass accretion rate directly and not by an
\mdot-induced environmental change.

The behavior of the bursts, in particular the burst oscillations, in
\sce\ is very rich and many questions remain for which we have no
adequate answers. Although the production mechanism behind the NCOs is
not well understood, their coherence and their frequency stability over
several years strongly suggest that indeed the oscillations are due to
the spin frequency of the neutron star. But it is still difficult to
imagine how the NCOs in the tail of bursts are formed. The entire
surface of the neutron star is expected to be involved in the
thermonuclear flash after the first few seconds. In such an environment
a hot spot would probably not survive. A possible solution to this
quandary is to invoke a different mechanism to produce the NCOs in the
decay phase of bursts. However, burst 17 in our sample (see
Fig.~\ref{zns}) shows what appears to be a continuous evolution of the
NCO frequency. This strongly suggest that only one mechanism is behind
the NCOs.

When we observe the oscillations in the tail of the burst, we usually
also see an increase in the frequency by a few Hertz as the burst
progresses \citep[see also][]{oscdisc,stroh98,zstroh}.  
The observed evolution of the oscillation frequency can be convincingly
attributed to the expansion of the neutron star atmosphere and
conservation of angular momentum (see \S~\ref{intro}). However, for at
least one of our bursts (burst 17, see Fig.~\ref{zns}) we observe an
increase in frequency of about 3.5~Hz. This would mean that the burning
layer must have expanded by almost 50 meters. It is not clear if such a
large expansion can be accounted for by the expanding layer
interpretation, although the expansion can be at least as large as 40
meters \citep{cumming}. Such a large increase in frequency is not unique
to \sce.  An increase of 5~Hz was recently found in the new NCOs source
X~1658--298 \citep{disc1659}, which is even larger than what we have
found for \sce.  However, because of the higher frequency of the NCOs in
X~1658--298 ($\sim$567~Hz) compared to that in \sce, the burning layer
has to have expanded more in \sce~than in X~1658--298 (50 meters versus
40--45 meters).

When we investigated the strength of the oscillations in detail, we
found that it is well correlated with inferred mass accretion rate: the
oscillations increase in strength when the accretion rate of the system
increases.  This correlation is more evident within bursts observed only
a few weeks apart, and becomes less clear for bursts separated in time
by several months to years.  It is unclear why the correlation is strong
on the short term (weeks) but fails to hold on longer timescales (months
to years).  Whatever the modulation mechanism, it will likely depend on
the state of the fuel layers on the neutron star, which may change
considerably with long term trends in the mass accretion rate. If we
think of the state of the neutron star surface as determining the moment
for the onset of the modulation mechanism, then one can imagine that the
overall changes in \mdot\ over months and years are resurfacing the
neutron star and hence the ``critical'' moment will be reached at
slightly different accretion rates on subsequent trips down the atoll
track. In other words, if the time averaged accretion rate has been very
low for a long interval of time, the structure of the neutron star
surface might be different than when the time averaged accretion rate
was considerably larger. 

The influences of the mass accretion rate on the burst oscillation
properties strongly suggest that they are indeed manifestations of
physical properties of the atoll systems.  We would expect that this
behavior of the NCOs will be also present in other sources. However, no
detailed studies have been performed for the other sources similar to
our study of \sce\ and that of KS~1731--260 by \citet{muno}. It is
important to determine whether the same correlation will hold for all
the sources which have NCOs.  However, we can definitely exclude the
possibility that all X-ray burst sources will show NCOs when they are at
high mass accretion rates. For at least two sources (Serpens X-1 and 4U
1705--44), X-ray bursts have been observed when they were on the banana
branch, thus at relatively high mass accretion rates. No NCOs were
discovered during those bursts (R. Wijnands 2000, private communication;
\citealt{ford}).  A more intrinic reason (\eg, binary inclination, the
strength and/or orientation of the magnetic field) must be found to
explain the many burst sources that do not show NCOs. On the other hand,
it is still possible that a considerable fraction of those sources have
indeed only been observed in their island states and might still show
NCOs when they can be observed to burst in their banana branches.

To conclude it is clear that from the study of \sce\ and KS~1731--260
we can already glean some common behaviors and many differences. More
atoll systems must be searched for NCOs in their bursts to begin
differentiating between those trends that are universal accross atoll
sources and those that are determined by intrinsic differences of the
systems.

{\it Note added in manuscript}.  
When we were about to submit our paper we became aware of the paper in
preparation by \citet{vs00} presenting a similar analysis of
\sce. Although their study is limited to the AO1 and AO2 observations,
they confirm our main conclusions: the NCOs are only present during
bursts at the highest inferred mass accretion rates and the strength of
the NCOs increases with accretion rate. They distinguish between two
types of PRE episodes, those marked by a small increase in radius after
the main PRE episode, which they termed {\it unusual}, and the more
standard continuous radius decrease after the PRE episode. They find
that standard PRE bursts only show NCOs after the peak of the burst and
unusual PRE episodes occur only in bursts at the lowest \mdot. We note
that these two correlations also hold when including our AO3 data set.

\acknowledgements

I thank Craig Markwardt for extremely helpful and well documented
software, Angela Olinto for unwavering support, Rudy Wijnands for
several critical readings of the manuscript and Tod Strohmayer for
unselfishly donating his data, time and knowledge and for teaching me
all about RXTE and \lmxb s. I thank Steve van Straaten and collaborators
for their willingness to exchange manuscripts before publication.  This
work made use of public RXTE data obtained through the GSFC HEASARC
archive, and was supported in part by NASA through grant
NRA-99-01-ADP-195, by NSF through grant AST-0071235 by DOE through grant
DE-FG0291.

\end{document}